\documentclass[aps,pre,twocolumn,groupedaddress]{revtex4-1}

\usepackage{graphicx}
\usepackage{ifpdf}
\ifpdf
  \DeclareGraphicsExtensions{.pdf,.png,.jpg}
\else
  \DeclareGraphicsExtensions{.eps}
\fi
\usepackage{amsmath}
\usepackage{amssymb}
\def\*#1{\mathbf{#1}}

\begin{document}


\title{Inferring the contiguity matrix for spatial autoregressive analysis
with applications to house price prediction}


\author{Somwrita Sarkar}
\email{somwrita.sarkar@sydney.edu.au}
\thanks{corresponding author}
\affiliation{Design Lab, Faculty of Architecture, Design and Planning, University of Sydney, Australia NSW 2006}
\affiliation{ARC Centre of Excellence for Integrative Brain Function}

\author{Sanjay Chawla}
\affiliation{Faculty of Engineering and IT, University of Sydney, Australia NSW 2006}
\affiliation{QCRI, Qatar}


\date{\today}

\begin{abstract}
Inference methods in traditional statistics, machine learning and data mining 
assume that data is generated from an independent and identically 
distributed (iid) process. Spatial data exhibits behavior for which
the iid assumption must be relaxed. For example, the standard approach in 
spatial regression is to assume the existence of a contiguity
matrix which captures the spatial autoregressive properties of the data.
However all spatial methods, till now, have assumed that the contiguity
matrix is given apriori or can be estimated by using a spatial
similarity function. In this paper we propose a convex optimization formulation to solve the
spatial autoregressive regression (SAR)  model in which both the contiguity
matrix and the non-spatial regression parameters are unknown and inferred
from the data. We solve the problem using the alternating direction
method of multipliers (ADMM) which provides a solution which is both
robust and efficient. While our approach is general we use data from
housing markets of Boston and Sydney to both guide the analysis and
validate our results. A novel side effect of our approach is the
automatic discovery of spatial clusters which translate to 
submarkets in the housing data sets.

\end{abstract}


\maketitle

\section{Introduction} 
\label{sec1}
The inferential building block of conventional statistics, data mining and machine learning rests on the assumption that data is
identically and independently distributed (iid). For example, the central limit theorem that the average of random variables (with 
finite variance) approaches a Normal distribution assumes the random variables are iid. Similarly standard linear regression postulates
a relationship between a dependent and independent variables as
\begin{equation}
\mathbf{y} =  \mathbf{X}\mathbf{\beta} + \mathbf{\epsilon},
\end{equation}
where $\mathbf{y}$ is the dependent variable, $\mathbf{X}$ is the matrix in which each column represents an independent variable, and $\mathbf{\epsilon}$ is the error vector, assumes that the errors ($\mathbf{\epsilon}$) are iid. The regression parameters $\beta$ when estimated then describe the dependent variable as a linear combination of the independent variables. 

A recurring problem in spatial statistics and data mining has been to carry out a spatial inferential task without invoking the iid
assumption. Indeed spatially indexed random variables neither tend to be independent nor identical. In fact Tobler's first law of geography~\cite{tobler}
that {\em Everything is related to everything else, but things that are nearby are more related than distant things} is clearly a statement against 
making an iid assumption in the analysis of spatial data. A common approach has been to
extend the inferential process by using a spatially autoregressive step\cite{anselin2012new,lesage2008introduction,shekhar2003spatial}
\begin{equation}
\mathbf{y} =  \rho\mathbf{W}\mathbf{y} + \mathbf{X}\mathbf{\beta} + \mathbf{\epsilon}
\end{equation}
Here, if on performing standard linear regression the residual error term $\mathbf{\epsilon}$ shows spatial dependence, then $\mathbf{W}$ the spatial contiguity or adjacency matrix is introduced to capture all the spatial dependency present in the data. The value of the parameter $\rho$ is estimated (between 0 and 1) along with the non-spatial parameters $\beta$, and if it is large (towards 1), then this is taken as evidence of spatial effects in the data. A large body of work has developed to exploit the above model for spatial regression tasks. 

However, two specific problems are identified in the literature. First, $\mathbf{W}$, which could take many different forms based on local connectivity assumptions, is always assumed a priori and imposed upon the problem~\cite{anselin1988, lesage2009}. Spatial dependence of the non-spatial independent or dependent variables, as we will see, could take complex forms, and an a priori assumption of $\mathbf{W}$ could at best hide latent dependences (by not modeling long range spatial dependences since standard forms of $\mathbf{W}$ model only local spatial dependence, for example), or at worst introduce artifact dependences (by positing that two points close in space are similar even if they have very different behaviors, for example). The role of this $\mathbf{W}$ has been a source of much debate in the econometrics community~\cite{lesage2010, gibbons2012, corrado2012}, and its modeling is acknowledged as an open problem. 

Second, even when the spatial parameter $\rho$ is estimated, it is unclear how $\mathbf{W}$ itself could be helpful in making predictions on the nature of the non-spatial dependent or independent variables, beyond demonstrating that a particular problem has strong spatial effects. If however, the $\mathbf{W}$ were estimated from the data along with the other regression parameters, it could be used to understand the spatial dependence inherent in the problem and then used to predict dependences and clustering properties of either the dependent or the independent variables. It could also implicitly explain the spatial effects that come into play through omitted variables that have not been modeled. 

In addressing the above issues, in this work we propose a convex optimization formulation of the problem where we infer both the contiguity matrix $\mathbf{W}$ and the 
regression parameters $\beta$~\cite{boyd2004convex}. Inferring both $\mathbf{W}$ and $\beta$ has several advantages in capturing the specifics of the problem that are not captured in a standard assumption of $\mathbf{W}$:
\begin{enumerate}
\item By inferring $\mathbf{W}$ from the data we are directly able to infer both short distance
connections and long distance connections. For example, houses along a major transport link (rail or highway) in a city may show correlations in prices over both short and long distances. Usually, the a priori assumption of $\mathbf{W}$ captures only short range connectivity and misses out on the long range connectivity properties that may exist in the data. 
\item By inferring $\mathbf{W}$ from the data we are directly able to capture clustering properties. Usually, the a priori assumption of $\mathbf{W}$ captures only local contiguity and misses out on clustered behavior that may exist in the data. This allows us to capture two different clustering effects: (i) a cluster that is discontinuous in space, (e.g., two housing submarkets in a city sharing dwelling stock and socio-economic characteristics may not be phyiscally adjacent but their
house prices maybe directly correlated), and (ii) different clusters that overlap in space, (e.g. low priced and high priced distinct submarket clusters may be superposed at the same point in space.)
\end{enumerate}

However their are certain unique challenges that emerge when dealing with the above problem:
\begin{enumerate}
\item The number of data items will always be less than the number of parameters. For $n$ locations in space and $p$ explanatory variables, the task is to estimate $n(n-1) + p$ parameters. We address this by a sparsity regularization condition on $\mathbf{W}$, in which we choose an optimum with minimum $L1$ norm. Such a sparse solution also causes the ``picking out" of the most important spatial dependence relations. At the same time, it ensures that we choose an optimum that satisfies a given condition out of the infinite possible optima. 
\item How should one carry out a validation of $\mathbf{W}$; i.e., once we estimate a $\mathbf{W}$, how do we validate whether its learnt form is actually giving us meaningful information? We address this by posing the optimization formulation in a form that ensures consistency with Ordinary Least Squares (OLS) solutions. That is, when the sparsity condition on $\mathbf{W}$ is removed, we obtain the non-spatial least squares solution, since $\mathbf{W}$ is forced to be close to zero.The OLS solution is treated as the baseline for prediction accuracy. Then, when the weight on the sparsity condition is increased, and we learn a sparse $\mathbf{W}$, then the prediction accuracy should improve. Additionally, using the Boston and Sydney housing market data, we also validate the cluster structure detected by $\mathbf{W}$ by comparing it with actual price submarkets in the data, to ascertain that the spatial dependence of the dependent variable being uncovered by $\mathbf{W}$ is meaningful. 
\item Finally how can the inferred parameters be used in a given application setting? We address this by showing that the learnt $\mathbf{W}$ leads to both better accuracy of prediction as well as the simultaneous identification of cluster structure in the data. In the context of housing markets, this leads to simultaneous estimation of global regression parameters along with the identification of housing submarkets and spatial spillover effects. Once the submarkets have been identified, independent local OLS could be run for these to identify subsets of explanatory variables that play special roles in specific submarkets. 
\end{enumerate}

The rest of the paper is organized as follows. Section~\ref{background} provides a background on the application setting: housing markets. Section~\ref{probdef} formulates the optimization problem, and section~\ref{algo} provides the ADMM solution and algorithm. Section~\ref{results} presents the results of applying the model and algorithm to housing market data from Boston and Sydney. Finally, related work is presented in section~\ref{relatedwork} and we end with conclusions in section~\ref{conclusions}. 

\section{Background: Housing markets as an application setting}
\label{background}

The approach we present in the paper is completely general and can be applied wherever spatial data is used. However, for concreteness, all our examples and experimental results will be based on housing markets analysis and predictions. 

The primary tool in housing econometrics for predicting relationships between a dependent variable (e.g. price) and a set of independent variables (e.g. socio-economic or hedonic variables) is regression. Traditional regression studies are modeled, however, on the assumption that the entire city or region is a unitary housing market~\cite{jones2009, watkins2001}. If the entire city or region is a single market, then traditional regression models need not consider spatial dependence beyond global location variables. The parameters of regression could then predict price and other trends in terms of sets of socio-economic and hedonic variables that operate consistently for the entire market. However, as new research reveals, this view is problematic. 

The current mainstream view of a unitary housing market stems from the seminal work on access-space models by Alonso and Muth~\cite{alonso1964, muth1969}, where the price gradient for housing costs is modeled in terms of the travel distance at which housing is located from the Central Business District (CBD). Although powerful in terms of providing analytically tractable long term equilibrium behaviors in terms tradeoffs of location, travel costs, and housing price, these models make a number of simplifying assumptions. There are no distinctions of housing or dwelling types, the city is embedded in a featureless space, travel costs are the same in all directions, and a monocentric city is assumed with a single CBD. These assumptions, however, do not hold for most cities. Thus, while the access-space models explain the large scale suburbanisation behaviors successfully, they are unable to account for the heterogeneity of the housing market, observed by the consistent presence of housing submarkets within a larger market. In the presence of polycentricity of urban structure (as opposed to one CBD) and other socio-economic, transport, land use and housing stock characteristics that break the symmetry of the monocentric price gradient model, a single city or region's housing market could show price gradients of a highly variable nature. A significant body of research as well as current market conditions now provide convincing evidence of quasi-independent housing submarkets within a highly segmented unitary housing market~\cite{bourassa1999, watkins2001, jones2005, jones2009, randolph2005, randolph2013, monashReport2013, domain2015}.
 

A housing submarket is defined as a set of dwellings that are close substitutes for other dwellings in the same submarket, but a poor substitute for dwellings in other submarkets. In terms of spatial distribution, views are divided between whether such submarkets are geographically defined, or defined on the basis of structural variables such as dwellings types. While the geographical definition provides spatially demarcated and non-overlapping submarkets, the structural definitions provide spatially segregated and discontinuous submarkets scattered through the city (e.g. as has been defined for Sydney and Melbourne~\cite{randolph2013}. However, it is accepted that neither the geographical nor the structural definition by itself can provide accurate identification of submarket structures, and that both georgaphy and structural features must be considered jointly for the task~\cite{watkins2001}. 

To the best of our knowledge, we do not know of algorithms, methods or theory that addresses how geographical heterogenous clustering and global structural trends can be brought together jointly to accurately identify housing submarkets. Even though the regression and clustering problems are intimately related, current approaches are usually sequential and either the regression problem sits divorced from the clustering problem, or the clustering problem is qualitatively pre-treated in ad-hoc ways~\cite{jones2005, watkins2001, bourassa1999}. This concrete application problem also outlines the general need for learning the structure of short range, long range, or clustered spatial dependence along with the simultanoeus estimation of regression parameters.

\section{Problem Definition}
\label{probdef}  

Let $\mathbf{y}$ be an $n \times 1$ vector of prices (or any other dependent variable), let $\mathbf{X}$ be an $n \times p$ matrix, each column of which represent one of $p$ explanatory variables, and let $\beta$ be a $p \times 1$ vector of regression parameters. Then the standard regression model is stated as $\mathbf{y} = \mathbf{X}\beta + \mathbf{\epsilon}$, where $\mathbf{\epsilon}$ represents the error term. In the presence of spatial autocorrelation, the error term does not show the identically independently distributed (i.i.d.) structure, but instead shows spatial patterns of dependence. 

An example standard response in spatial econometrics and statistics is a Spatial Autoregressive Model (SAR)  model of the form: $\mathbf{y} = \rho \mathbf{Wy} + \mathbf{X} \beta + \mathbf{\epsilon}$, where the matrix $\mathbf{W}$ is an $n \times n$ spatial dependence matrix. As discussed, this $\mathbf{W}$ is modeled based on local distance, contiguity or lattice type assumptions. The principle observation is that it is imposed onto the problem as a data term, instead of being treated as a variable. 

We now consider the more realistic case where we treat $\mathbf{W}$ as a variable and learnt. In the case of housing prices, particular neighborhoods for example show similar price trends because of the common presence of highly regarded local amenities such as school quality or negative criteria such as crime rates. Thus, it is likely that the price data will show spatial clustering, in which case $\mathbf{W}$ could have a clustered structure with short and long range dependences rather than following local lattice, distance or contiguity assumptions. In other words, $\mathbf{W}$ could have a higher-order structure that could be revealed using the data itself. 

Since we have to estimate $n(n-1) + p$ parameters, and the number of variables is always going to be higher than the data, we will have an infinite number of solutions. In this instance, we draw upon the condition that $\mathbf{W}$ has to be sparse and positive, since a dense $\mathbf{W}$ implies every location is related to every other, and is unlikely to provide meaningful information. Hence we learn the $\mathbf{W}$ with the minimum $L1$-norm.     

Thus, the optimization problem can be stated as: \begin{equation} 
\label{optProb1} 
\begin{split} 
  \mbox{min       } \frac{1}{2} ||\mathbf{y} - \mathbf{Wy} - \mathbf{X} \beta ||_{2}^{2} + \lambda_{1} ||\mathbf{W}||_{1}  \\ 
  \mbox{sub to       } diag(\mathbf{W}) = 0, \\ 
  \mathbf{W} \geq 0
\end{split} 
\end{equation} 

\section{Optimization Algorithm}
\label{algo}
We will use ADMM~\cite{boyd2011distributed,parikh2014proximal} to solve the above optimization problem.  
Introducing auxiliary variable $\mathbf{A} \in \mathbb{R}^{n \times n}$, we can rewrite the above as \begin{equation} 
\label{finalModel} 
\begin{split} 
  \mbox{min       } \frac{1}{2} ||\mathbf{y} - \mathbf{Wy} - \mathbf{X}\beta||_{2}^{2} + \lambda_{1} ||\\mathbf{A}||_{1} \\ 
  \mbox{sub to       } \mathbf{A} = \mathbf{W} - diag(\mathbf{W}), \\ 
  \mathbf{W} \geq 0
\end{split} 
\end{equation} 

Then, the augmented Lagrangian is written as \begin{equation} 
\begin{aligned} 
L[\mathbf{W}, \mathbf{A}, \beta, \mathbf{\Delta_{1}}] = \\ 
& \frac{1}{2} ||\mathbf{y} - \mathbf{Wy} - \mathbf{X}\beta||_{2}^{2} ~+~ \lambda_{1} ||\mathbf{A}||_{1} \\
& ~+~ \text{tr}[\mathbf{\Delta_{1}}^{T} (\mathbf{A} - \mathbf{W} + diag(\mathbf{W}))] \\
& ~+~ \frac{\rho_{1}}{2} ||\mathbf{A} - \mathbf{W} + diag(\mathbf{W})||_{2}^{2} ~+  I_{+}(\mathbf{W})
\end{aligned} 
\end{equation} 
where $ I_{+}(\mathbf{W}) = 0$ when $\mathbf{W} \geq 0$ and $\infty$ otherwise.

\subsection{Updating $\mathbf{W}$}

\begin{equation} 
\begin{aligned}
\frac{\partial L}{\partial \mathbf{W}} = \frac{\partial}{\partial \mathbf{W}} \bigg [\frac{1}{2} ||\mathbf{y} - \mathbf{Wy} - \mathbf{X}\beta||_{2}^{2} ~+ \\
\text{tr}[\mathbf{\Delta_{1}}^{T} (\mathbf{A} - \mathbf{W} + diag(\mathbf{W}))] ~+ \\
 \frac{\rho_{1}}{2} ||\mathbf{A} - \mathbf{W} + diag(\mathbf{W})||_{2}^{2} ~+  I_{+}(\mathbf{W}) \bigg ]
\end{aligned}
\end{equation}

Looking at the individual terms, 

\begin{equation}
\begin{aligned}
\frac{\partial}{\partial \mathbf{W}} \bigg [\frac{1}{2} ||\mathbf{y} - \mathbf{Wy} - \mathbf{X}\beta||_{2}^{2} \bigg ] \\
= \frac{1}{2} \frac{\partial}{\partial \mathbf{W}} \bigg [ \text{tr} (\mathbf{y} - \mathbf{Wy} - \mathbf{X} \beta) (\mathbf{y} - \mathbf{Wy} - \mathbf{X} \beta)^{T} \bigg ] \\
= \frac{1}{2} \bigg ( \frac{\partial}{\partial \mathbf{W}} tr \bigg [ -\mathbf{yy}^{T}\mathbf{W}^{T} - \mathbf{Wyy}^{T} + \mathbf{Wyy}^{T}\mathbf{W}^{T} \\
+ \mathbf{Wy}\beta^{T}\mathbf{X}^{T} + \mathbf{X}\beta \mathbf{y}^{T}\mathbf{W}^{T} \beta \bigg ] \bigg ) \\
= -\mathbf{yy}^{T} + \mathbf{Wyy}^{T} + \mathbf{X}\beta \mathbf{y}^{T}
\end{aligned}
\end{equation}

Similarly, 

\begin{equation}
\begin{aligned}
\frac{\partial}{\partial \mathbf{W}} \text{tr} [\mathbf{\Delta_{1}}^{T} (\mathbf{A} - \mathbf{W} + diag(\mathbf{W}))] \\
= - \mathbf{\Delta_{1}} + \text{diag}(\mathbf{\Delta_{1}})
\end{aligned}
\end{equation}

Similarly,

\begin{equation}
\begin{aligned}
\frac{\partial}{\partial \mathbf{W}} \frac{\rho_{1}}{2} ||\mathbf{A} - \mathbf{W} + diag(\mathbf{W})||_{2}^{2} \\
= \rho_{1} (-\mathbf{A} + diag(\mathbf{A}) + \mathbf{W})
\end{aligned}
\end{equation}

Combining all the individual terms and setting them to zero, and applying the positivity constraint on $\mathbf{W}$, we get 


\begin{equation}
\begin{aligned}
\mathbf{W}^{*} = \bigg [ [\mathbf{yy}^{T} + \rho_{1} \mathbf{I}] ^ {-1}
(\mathbf{yy}^{T} - \mathbf{X}\beta \mathbf{y}^{T} + \mathbf{\Delta_{1}} - \text{diag}(\mathbf{\Delta_{1}}) + \\
\rho_{1} \mathbf{A} + \rho_{1} \text{diag} (\mathbf{A}) ) \bigg ]_+
\end{aligned}
\end{equation}

\subsection{Updating $\mathbf{A}$}

\begin{equation}
\begin{aligned}
\frac{\partial L}{\partial \mathbf{A}} = \frac{\partial}{\partial \mathbf{A}} \bigg [ \lambda_{1} ||\mathbf{A}||_{1} + \text{tr}[\mathbf{\Delta_{1}}^{T} (\mathbf{A} - \mathbf{W} + diag(\mathbf{W}))] \\
+ \frac{\rho_{1}}{2} ||\mathbf{A} - \mathbf{W} + diag(\mathbf{W})||_{2}^{2} \bigg ]
\end{aligned}
\end{equation}

Applying the soft-thresholding operator, 

\begin{equation}
S_{f}(x) = \text{sign}(x)*\text{max}(|x|-f, 0)
\end{equation}

we will get

\begin{equation}
\begin{aligned}
\mathbf{A}^{*} = \mathbf{C} - diag(\mathbf{C}), \\
\mathbf{C}= S_{\frac{\lambda_{1}}{\rho_{1}}} \bigg ( \mathbf{W} + \frac{\mathbf{\Delta_{1}}}{\rho_{1}} \bigg ).
\end{aligned}
\end{equation}

%

\subsection{Updating $\beta$}

\begin{equation}
\begin{aligned}
\frac{\partial L}{\partial \beta} = \frac{\partial}{\partial \beta} \bigg [\frac{1}{2} ||\mathbf{y} - \mathbf{Wy} - \mathbf{X}\beta||_{2}^{2} \bigg ]
\end{aligned}
\end{equation}

This can be solved as 
\begin{equation}
\begin{aligned}
\frac{\partial}{\partial \beta} \bigg [\frac{1}{2} ||\mathbf{y} - \mathbf{Wy} - \mathbf{X}\beta||_{2}^{2} \bigg ] \\
= \frac{1}{2} \frac{\partial}{\partial \beta} \bigg [ \text{tr} (\mathbf{y} - \mathbf{Wy} - \mathbf{X} \beta) (\mathbf{y} - \mathbf{Wy} - \mathbf{X} \beta)^{T} \bigg ] \\
= \frac{1}{2} \bigg ( \frac{\partial}{\partial \beta} \text{tr} \bigg [ -\mathbf{y} \beta^{T} \mathbf{X}^{T} + \mathbf{Wy} \beta^{T} \mathbf{X}^{T} - \mathbf{X} \beta \mathbf{y}^{T}   \\
+ \mathbf{X} \beta \mathbf{y}^{T} \mathbf{W}^{T} + \mathbf{X} \beta \beta^{T} \mathbf{X}^{T}  \bigg ] \bigg ) \\
= -\mathbf{X}^{T} \mathbf{y} + \mathbf{X}^{T} \mathbf{W y} + \mathbf{X}^{T} \mathbf{X} \beta )
\end{aligned}
\end{equation}

Setting this to zero, we get

\begin{equation}
\begin{aligned}
\beta^{*} = (\mathbf{X}^{T}\mathbf{X})^{-1} [\mathbf{X}^{T} \mathbf{y} - \mathbf{X}^{T} \mathbf{W y}]
\end{aligned}
\end{equation}

\subsection{Updating $\mathbf{\Delta_{1}}$}

Finally, we can update the Lagrangian multiplier $\mathbf{\Delta_{1}}$ as

\begin{equation}
\begin{aligned}
\mathbf{\Delta_{1}}^{*} = \mathbf{\Delta_{1}} + \rho_{1}(\mathbf{W}-\mathbf{A}) \\
\end{aligned}
\end{equation}

\textbf{Convergence.} This is run iteratively to convergence. Since this optimization problem is convex, ADMM is guaranteed to attain a global minimum. We have used a stopping criterion where we measure the absolute difference between the primal and dual variables and stop when the error is stabilized below a given threshold.

\section{Results}
\label{results}



We test our approach on two data sets: (a) the Boston housing data set, that provides the prices and explanatory variable measurements for 506 houses along with the latitude and longitude locations, and (b) a data set for 51 Local Government Areas (LGAs) of Sydney, further subdivided into 862 suburbs, with 2015 prices of houses and related explanatory variables available for both area definitions. 

We have independently run the optimization model on CVX as well as our ADMM algorithm, and while the results from both are the same, the CVX takes much more time as the number of locations ($n$) climbs even upto 500. For example, testing the $n=862$ Sydney suburbs dataset took about 960 seconds (about 16 minutes) on CVX, whereas it takes negligible time with ADMM. The inference of the spatial weights matrix $\mathbf{W}$ should, as one of its main contributions, scale to identifying fine scale spatial clustering (in this case, housing submarkets) for latitude-longitude level microdata upto the individual house level. This would mean big data sets with millions of locations. Hence, we establish the justification for proposing the ADMM algorithm. Our results, contributions and validations for each data set are reported in three sections, discussed below.

\subsection{Co-estimation of regression parameters and spatial dependence}
 First, we present a comparison of the result of our model with the ordinary least squares (OLS) and spatial autoregressive models (SAR), showing consistent estimates of $\beta$ across all the three cases.  However, in our approach, we also learn the structure of spatial dependence, through the spatial weights matrix $\mathbf{W}$, that is ignored in OLS and the form of which is assumed \textit{a priori} as a data term in SAR. While we use the SAR model only as one of the possible spatial regression models, we note that almost all spatial regression models assume an a priori fixed $\mathbf{W}$. This is problematic, because a priori assumptions rest on modeling only local spatial connectivity, whereas the data may contain a mix of short range and long range connectivity as well as clustering. Hence, to learn the structure of $\mathbf{W}$ would have many advantages. The main contribution of the approach presented in this paper is learning this $W$ along with the consistent estimation of the regression parameters.

We note here that the consistent estimation of regression parameters for the entire region is important, since regression parameters are required to be ``global" for most policy settings. Thus, a validation check on any identified structure of $\mathbf{W}$ is that its co-inference with $\beta$ does not lead to wildly deviant or fluctuating forms for $\beta$ and is largely consistent with OLS results. For example, even though in principle, it is possible to run local regressions for each LGA or census tract or suburb within a metropolitan or state region, policy decisions are usually made by authorities at the metropolitan or state level on the basis of the global paramters. It is also at the metropolitan or state level that housing submarket clusters need to be identified for the entire urban region. This would require that regression parameters be globally estimated for entire regions (as they usually are). This approach provides an added global - local connection, since it does not preclude the possibility of carrying out local regression too. Once the estimation of local spatial clusters are performed using $\mathbf{W}$, it is always possible to run local regression models for the identified submarket clusters (that are hard to define otherwise, as noted in the background section). 

\subsubsection{Boston housing prices data}

The Boston housing data set was produced in 1978, incorporating measurements of 14 variables across 506 census tracts~\cite{harrison1978}, and is available as part of the Matlab Spatial Econometrics package~\cite{lesage2009}. The log of median value of owner-occupied homes in \$1000's is taken as the dependent variable, and all the others as explanatory (per capita crime rate by town, proportion of residential land zoned for lots over 25,000 sq.ft., proportion of non-retail business acres per town, a Charles River dummy variable (= 1 if tract bounds river; 0 otherwise), nitric oxides concentration (parts per 10 million), average number of rooms per dwelling, proportion of owner-occupied units built prior to 1940, weighted distances to five Boston employment centres, index of accessibility to radial highways, full-value property-tax rate per \$10,000, pupil-teacher ratio by town, the proportion of black population by town, and lower status of the population).

\begin{figure}
\centering
\includegraphics[height = 3.9in, width = 2.8in]{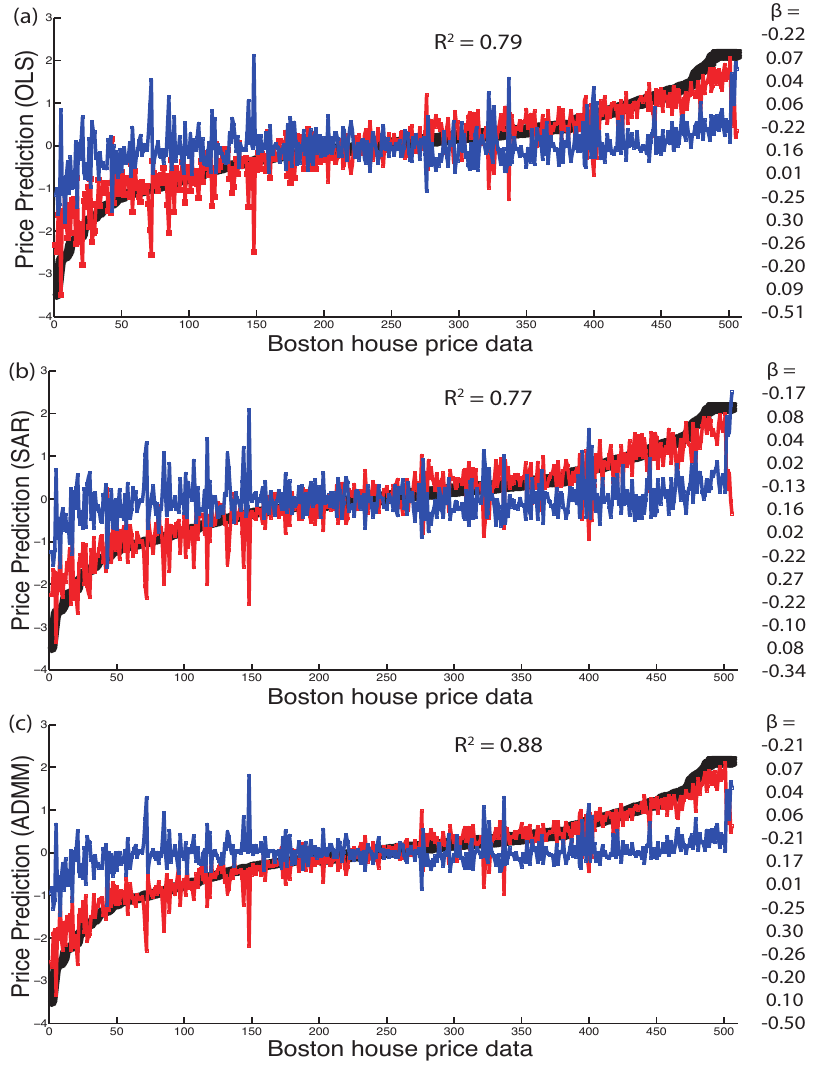}
\vspace{-0.5cm}
\caption{\label{bostonFig1}Boston house prices regression estimates. (a) OLS. (b) SAR. (c) ADMM, $\lambda_{1}=0.8$. The fit is better for ADMM with similar $\beta$ estimates. Range of $\lambda_{1}$ varied from 0 to 5. Black: actual prices, red: prediction, blue: residuals.} 
\end{figure}

We first report on the estimation of the regression parameters $\beta$. Figure~\ref{bostonFig1} shows the prediction from OLS, SAR, and ADMM, respectively, along with the $\beta$ estimates and corresponding $R^{2}$ values. The spatial weights matrix $\mathbf{W}$ for the SAR estimation was computed using the inbuilt routines in the Matlab Spatial Econometrics package. It is seen that the co-estimation of $\mathbf{W}$ causes the $R^{2}$ to rise, at the same time, maintaining the consistency of the regression parameters $\beta$. In the next section we report on analysing the structure of the inferred $\mathbf{W}$, but simply note for now that the co-estimation of spatial dependence from data can actually improve prediction capability, through the higher $R^{2}$ that results while holding $\beta$ consistent. 

\subsubsection{Sydney housing prices data}

At the time of writing this paper, the housing market in Sydney is witnessing a particularly turbulent time~\cite{smh2015, domain2015}. In 2014-15, there were warnings of an impending ``housing bubble", prices have risen consistently and steeply in the last few years, and housing is deemed as acutely unaffordable especially for the younger and poorer sections of the population~\cite{randolph2005, randolph2013}. By 2016, there are now warnings of an impending fall in prices. It is an important problem to estimate the fine scale spatial organization of housing submarkets in especially Sydney and to some extent Melbourne, since they are the regions witnessing the highest rises in price as well as demand population. To the best of our knowledge, no statistically or quantitatively based identification of Sydney and Melbourne housing submarkets have been performed to date, though there is ample evidence of policy interest in the issue and qualitative urban planning approaches that are directed towards tackling this problem~\cite{randolph2005, randolph2013}. The Australian Bureau of Statistics publishes capital city housing price index series, but assumes the entire metropolitan region as a single housing market, without the identification of clustered submarkets within the region that are witnessed and experienced by residents. Large parts of cities being unaffordable and out of reach pose serious socio-economic affordability and equity concerns. 

The agency that records all of the residential property transactions is Core Logic / R P Data. They have made the data available for University and academic research through an organization called SIRCA. We have obtained 9 months (latest at the time of writing this paper) of 2015 residential property transactions data for the Sydney housing market, that we aggregate over each geographical area of analysis to produce an annual aggregate. The metropolitan region of Sydney is divided into 43 Local Government Areas (LGAs) and about 8 outlying LGAs that are adjacent to metropolitan Sydney and seeing spillover growth. We have included all these 51 regions in our analysis. The 51 LGAs further subdivide into about 862 susburbs (roughly corresponding to individual post-code areas), that forms the lowest level of area definitions for which aggregated data is available from RP Data. Data is available for both houses and units (apartments) separately, and in this paper we consider the individual detached houses data, since Sydney is pre-dominantly a detached dwelling houses market (though the apartment market is becoming increasingly important recently due to increased population pressures). 

The dependent variable is median price per location (LGA or suburb), the log of which has been considered for the suburb level data, and the explanatory variables are the numbers of houses sold per location, the total monetary value in dollars of the total number of properties sold, the number of properties sold under \$200,000, from \$200,000 - \$400,000, \$400,000 - \$600,000, \$600,000 - \$800,000, \$800,000 - \$1,000,000, \$1 million - \$2 million, and above \$2 million, the rental median for each area, the total number of dwellings listed for selling, and the total numbers of dwellings in the area. 

\begin{figure}
\centering
\includegraphics[height = 3.1in, width = 3.2in]{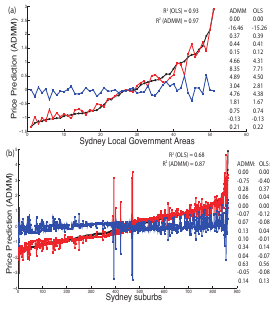}
\vspace{-0.5cm}
\caption{\label{Sydney_fig3}Sydney house prices regression estimates. (a) ADMM for 51 LGAs, $\lambda_{1}$ = 0.2 (b) ADMM for 862 suburbs, $\lambda_{1}=4$. The fit is better for ADMM with similar $\beta$ estimates. Range of $\lambda_{1}$ varied from 0 to 10. OLS estimates and $R^{2}$ values are also reported.Black: actual prices, red: prediction, blue: residuals.} 
\end{figure}

Figure~\ref{Sydney_fig3} shows the predictions of regression parameters from both OLS and the ADMM (the predictions from SAR are close to these). Again, it is seen that with consistent $\beta$ emerging from all approaches, the $R^{2}$ values are higher for the ADMM, showing a better fit. It is also interesting to observe in the Sydney case, that while the explanatory variables that measure how many properties are being sold within given price ranges of \$200,000 brackets emerge as very important at the LGA level, they turn out to be less important at the suburb level. This is due to the nature of the data itself - where at the suburb level, many of them may be ``low activity" zones within a particular LGA, and numbers being sold and reported are much lower than the same at the LGA level (with many zero values). However, the estimates are consistent where the data themselves are similarly scaled at both geographic levels. For example, the second variable is the total value of the sales in dollars, and is consistent for both area definitions. The eleventh variable is the rental median for the area, and is consistent for both area definitions. 

The consistency of the ADMM algorithm estimating regression parameters consistent with OLS is not simply an empirical observation, but can also be observed from the ADMM model. When we set the regularizer $\lambda_{1}$ to zero, we get OLS estimates from the ADMM model. Since with $\lambda_{1} = 0$, the algorithm places no special weight to spatial dependence, the resulting $\mathbf{W}$ estimate coming from the data term $0.5 * ||(\mathbf{y} - \mathbf{Wy} - \mathbf{X \beta})||^{2}_{2}$ results in a dense $\mathbf{W}$ but with extremely small close to zero entries. Then, the $\mathbf{\beta}$ that is estimated are the same as or extremely close to those estimated in the OLS. When the $\lambda_{1}$ value is increased, a sparser $\mathbf{W}$ is returned but with higher valued entries. Then, the $\mathbf{\beta}$ shifts in a consistent way, along with producing a sparse $\mathbf{W}$ matrix that brings out the spatial dependence in the data. The values of $\lambda_{1}$ used for the reported results are shown in the figure captions in each case, but in general a larger value of $\lambda_{1}$ is needed for larger data sets. The consistency of the estimated $\beta$ and the improvement of the fit of prediction from our approach provides a first validation for the structure of the learnt $\mathbf{W}$. 

\subsection{Spatial dependence and clustering captured in $\mathbf{W}$}

We now report on the nature of spatial dependence and clustering that is learnt in the $\mathbf{W}$ matrix. The approach presented in the paper models $\mathbf{y} = \mathbf{Wy} + \mathbf{X \beta} + \mathbf{\epsilon}$, showing that $\mathbf{W}$ should capture ``price" submarkets, or more generally, spatial dependence of the dependent variable. Therefore, to provide a second validation to the structure of $\mathbf{W}$, we compare against the spatial structure of the actual price data and the clustering predictions made by performing a spectral clustering of the learnt $\mathbf{W}$. 

We note here that the same approach can be extended to model ``demand" or ``supply" submarkets too, or more generally, clustering and short and long range spatial dependence of the explanatory variables. For example, we could have a model $\mathbf{y} = \mathbf{WX}\mathbf{\beta} + \mathbf{\epsilon}$, which is not convex but bi-convex in $\mathbf{W}$ and $\mathbf{\beta}$, and would need a different ADMM algorithm to be derived. But, the general principle of deriving the spatial dependence will hold. 

Estimating $\mathbf{W}$ has a particular advantage that is motivated by an omitted variables observation. In any regression model, it is impossible to capture all the variables that affect the dependent variable, due to factors such as unknown variables or unavailable data. In such a case, estimating spatial dependence via $\mathbf{W}$ can be used to implicitly predict how changes in a particular location could lead to changes in all other locations without explicitly knowing about all the other variables that have been omitted. 

\subsubsection{Boston housing prices data}

\begin{figure*}
\includegraphics[height = 2.5in, width = 7in]{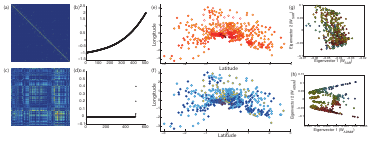}
\vspace{-0.5cm}
\caption{\label{bostonFig2} Structure of inferred $\mathbf{W}$ for Boston. (a) $\mathbf{W}$ SAR. (b) Eigenvalues of (a). (c) Learnt $\mathbf{W}$ ADMM. (d) Eigenvalues of (c). (e) Actual prices. (f) Derived clusters from $\mathbf{W}$. (g,h) Top two eigenvector components from SAR and ADMM $\mathbf{W}$s, respectively, colors represent prices.} 
\end{figure*}

Figures \ref{bostonFig2}(a-d) show the learnt $\mathbf{W}$ matrix using our algorithm, compared to the pre-fixed $\mathbf{W}$ used in SAR: instead of only local spatial connectivity the learnt $\mathbf{W}$ captures both long range connectivity between locations as well as clustering behavior. We test this finding by performing an eigenvalue decomposition of the learnt $\mathbf{W}$, identifying the number of leading eigenvalues $k$ separated from the bulk distribution of eigenvalues, and then clustering  lower $k$-dimensional representation of $W$ using K-means~\cite{SarkararXiv, Sarkar2013, Sarkar2011}. Such clustering is not identified for the $\mathbf{W}$ from SAR [Fig.~\ref{bostonFig2}(b)]. Figure~\ref{bostonFig2}(e) shows the real price distribution in the 506 locations, where blue corresponds to lowest prices and red to highest prices. Note that the spatial clustering is not local, and there are instances where lower priced houses occur in close proximity to higher priced houses. The clusters are therefore mixed or superposed in space. Moreover, there are also long range spatial dependences, with similar priced houses heterogenously spread across in space. Assuming purely local spatial dependence as spatial regression models assume will not capture these superpositions of clusters in short range or long range spatial dependence of prices as observed in the Boston data. Figure~\ref{bostonFig2}(f) shows the clusters returned by the Kmeans algorithm performed over the lower 4-dimensional representation of our learnt $\mathbf{W}$: the price clusters retrieved capture meaningful information; the highest priced cluster is yellow (white in Fig.~\ref{bostonFig2}(e)) , the second highest price cluster is light blue (yellow in Fig.~\ref{bostonFig2}(e)), the third highest price cluster is dark blue (organge in Fig.~\ref{bostonFig2}(e)), and the lowest priced one is dark red (dark red also in Fig.~\ref{bostonFig2}(e)). The derived clusters correlate strongly with price data as well as short range and long range spatial dependences.

To validate and test this finding more strongly, we plot the top two eigenvector components of both the SAR $\mathbf{W}$ and the one learnt from the ADMM algorithm. Figure~\ref{bostonFig2}(g,h) show the top two eigenvector components of the SAR based and ADMM based $\mathbf{W}$ matrices, respectively, with the eigenvector components colored by price data. As is clearly visible, the SAR based $\mathbf{W}$ does not capture any clustering or short or long range dependence, since by definition, it is defined purely on the basis of local connectivity between points. In contrast, the top two eigenvectors of the $\mathbf{W}$ derived using the ADMM algorithm shows clear clustering behavior.


\subsubsection{Sydney housing prices data}

\begin{figure*}
\centering
\includegraphics[height = 3.5in, width =7in]{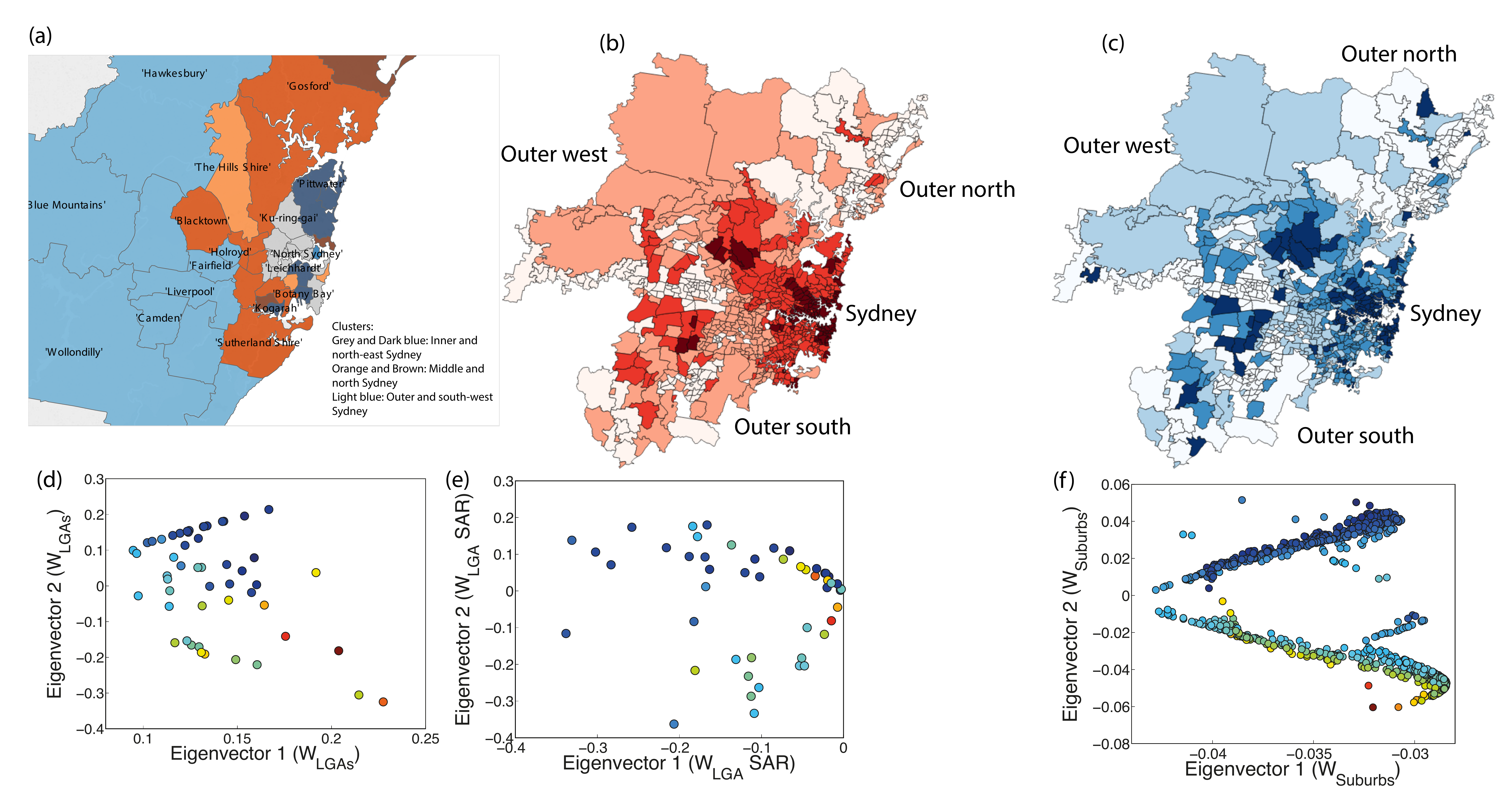}
\vspace{-0.5cm}
\caption{\label{LGA_Clusters}Structure of inferred $\mathbf{W}$ for Sydney. (a) Clusters from ADMM $\mathbf{W}$ at the LGA level. (b) Actual prices (c) Clusters from ADMM $\mathbf{W}$ at the suburb level. (d,e) Top two eigenvector components from SAR and ADMM $\mathbf{W}$s, respectively, at the LGA level, colors represent prices. (f) Top two eigenvector components from ADMM $\mathbf{W}$ at the suburb level, colors represent prices.} 
\end{figure*}

Figure~\ref{LGA_Clusters} shows the spatial structure of $\mathbf{W}$ derived using the ADMM analysis for the Sydney housing prices data. Again, just like the Boston example, we have performed a spectral clustering of the derived $\mathbf{W}$. At both the LGA and suburb levels, the identified clusters are spatially plotted using Geographic Information Systems (GIS) software to show the spatial nature of $\mathbf{W}$ along with the short range and long range dependencies derived from it, in addition to clustering behavior. We note here again that such analysis cannot be performed for the local contiguity matrices of the kind used in spatial regression models such as SAR. 

\begin{figure}
\centering
\includegraphics[height = 2.4in, width =3in]{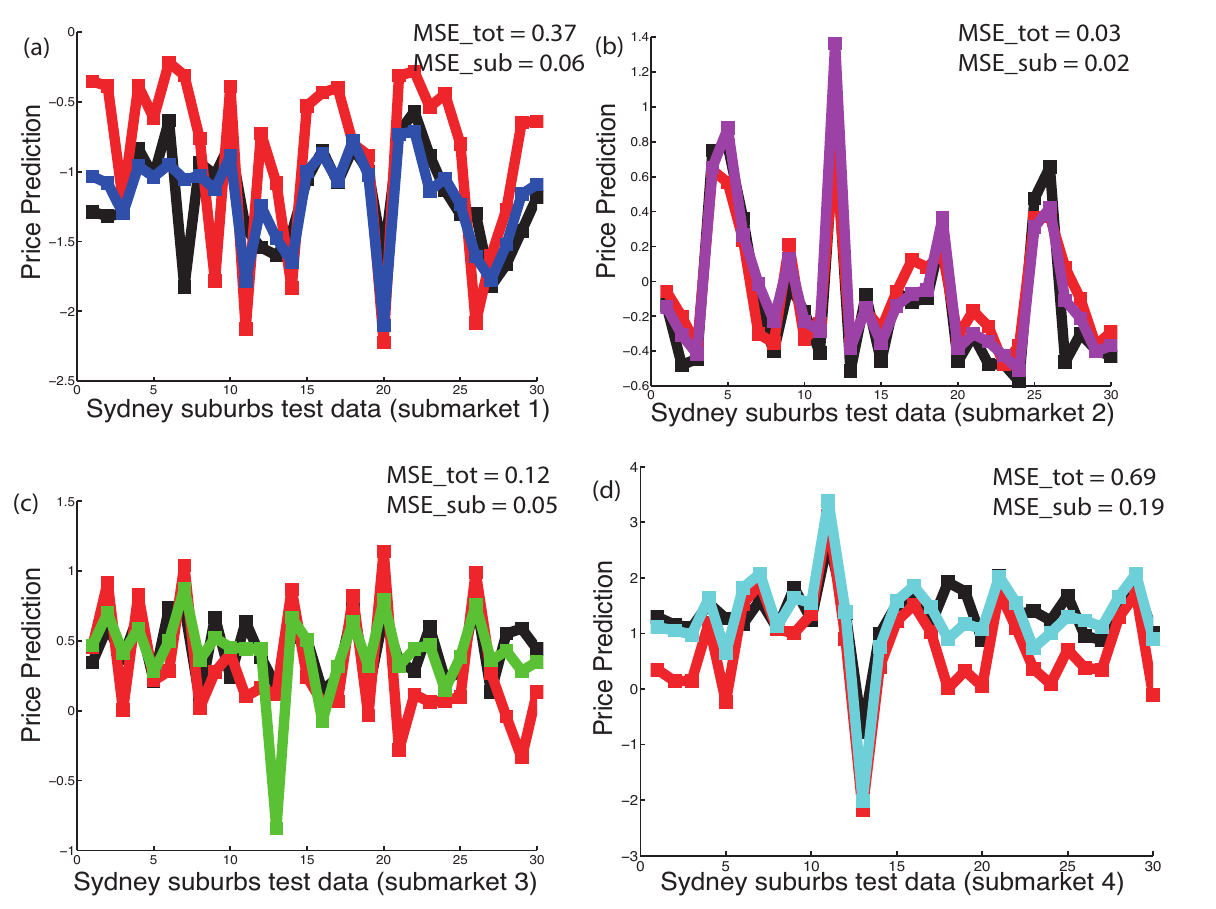}
\vspace{-0.5cm}
\caption{\label{submarkets} Local prediction for submarkets. (a-d) Price predictions for the Sydney suburbs test sets for the 4 identified submarkets using learnt $\mathbf{W}$. Black: original price, red: Whole market predictions, colored lines (blue, magenta, green, cyan): local market predictions.} 
\end{figure}

\begin{figure}
\centering
\includegraphics[height = 1.5in, width = 3in]{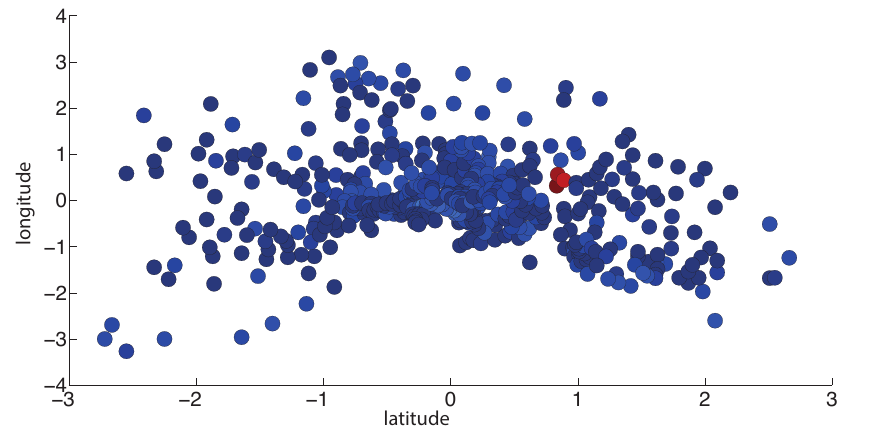}
\vspace{-0.5cm}
\caption{\label{bostonFig3}Illustration of spatial spilliover effects predicted using learnt $\mathbf{W}$} 
\end{figure}

Figure~\ref{LGA_Clusters}(a) shows the clusters identified for the 51 LGAs of the Sydney Metropolitan area. We interpret the clusters using our local knowledge of the housing market in Sydney. There were 7 clusters identified, corresponding to the 7 leading eigenvalues of $\mathbf{W}$. Sydney is a coastal city, with its Central Business District (CBD) at the eastern most end next to the sea. From the CBD, the city is roughy divided into inner, middle and outer rings of Sydney. The price structure roughly follows this spatial structure with the inner ring the most expensive and the outer ring the least expensive. However, there is also a very pronounced north-east (much more expensive) and south-west (less expensive) divide~\cite{randolph2005}. Figure ~\ref{LGA_Clusters}(a) shows that all of these effects have been identified in the clusters: the grey and dark blue colors represent the most expensive inner ring LGAs. Note that the more expensive LGAs are more pronounced towards the east and the north. The light and dark orange clusters roughly represent the middle ring. Again, the more expensive LGAs are stretched toward the north (e.g. the Hills Shire in the north colored light orange is a significant distance away from the CBD, but is very expensive with respect to the median and is placed in a cluster along with more distant LGAs closer to the city CBD). Finally, the least expensive LGAs are in the outer ring and show a pronounced south-west concentration. 

Figure~\ref{LGA_Clusters}(b) and (c) show the actual prices and the clusters identified for the 862 suburbs into which the 51 LGAs of the Sydney Metropolitan area are divided into. 4 main clusters were identified corresponding to the 4 leading eigenvalues of $\mathbf{W}$. Other small eigenvalues close to the bulk but just merging into the bulk exist, and it is possible that the clustering may have a hierarchically nested structure if more eigendimensions are chosen for the clustering~\cite{Sarkar2011, Sarkar2013}. We will explore this in future work. However, the spatial organization of actual prices [Fig.~\ref{LGA_Clusters}(b)] and the clusters identified through $\mathbf{W}$ [Fig.~\ref{LGA_Clusters}(c)] are clearly similar. One observation is that at the suburb level the clusters are much more fractured. The reason for this is that there is a lot of heterogeneity within single LGAs, with price and socio-economic differences between them. Thus, contiguous groups of suburbs belonging to different LGAs could belong to the same cluster at this finer suburb level. Second, even with the fractured picutre, there are major patterns visible: (a) the north-east suburbs of main Sydney city largely belong to different clusters as compared to the south west suburbs of main Sydney, but in 2015 pockets of high priced activity submarkets are visible towards the outer south and outer far north and (b) the highest priced suburbs (mostly orange red and dark to middle blues) around the city extend into the west and south along where the major transportation links are (railway, freeways etc.).


Figure~\ref{LGA_Clusters}(d) and (f) show the top two eigenvector components of $\mathbf{W}$ for the LGAs and suburbs, respectively. Again, the localization of eigenvector components by price is clearly visible. This clustering is not visible for a local contiguity matrix that we built by hand for the LGAs of Sydney, in which we connected each LGA to other LGAs with which it shares boundaries. The corresponding eigenvector components of this hand-crafted $\mathbf{W}$ or other similar variants based on local contiguity needed for running the SAR algorithm or other spatial regression models is shown in Fig.~\ref{LGA_Clusters}(e).

\subsection{Predictions over local submarkets and spatial spillovers} 

Once the structure of $\mathbf{W}$ has been learnt, and local price submarkets (or dependent variable clusters) identified, local regression can be performed for each of the submarkets. We have divided the entire data into the 4 identified submarkets, and then divided each of the 4 submarkets into training and test sets, with the test sets containing 30 data points each. This implies 120 randomly picked data points separated out from the whole data set of 862 suburbs. Then, we have performed OLS for the whole training set (i.e., $862-120=742$ data points in the test set), followed by OLS on the training sets for each of the individual submarkets (i.e., minus the 30 test data points from each of the submarkets). Then, we have used the test set of 120 data points to estimate the price prediction using the OLS estimates from training on the 742 suburbs. In parallel, we have predicted the price for the same 120 points, but this time divided into the 30 data points per submarket, and used the submarket OLS estimates to predict the local submarket price for these points. Figure~\ref{submarkets} shows that in each case, the local submarket prediction is more accurate than the the whole market prediction. This demonstrates that by using the structure of the learnt $\mathbf{W}$ better local regression estimates can be produced for specific housing submarkets. This is important from the policy perspective, because planning authorities could consider the regression parameter estimates from both the global whole city level as well as the local submarkets levels, to specifically inform spatial decisions such as new land release or density based rezoning.

Finally, we note that in spatial regression, a change in one of the explanatory variables could result in spatial spillover effects, such that the dependent variable is affected for several locations by a change in a single location. This is not considered in OLS, where a change in the explanatory variables changes the prediction for only one data point. Using both the estimated $\beta$ and $\mathbf{W}$, changes in the explanatory variables can be used to study the spatial spillovers for all the dependent variables. 

To predict the spatial spillover effect, we change the values of a few explanatory variables, producing a new $X_{new}$ and use the learnt $W$ and $\beta$ to recompute a new price vector $y_{new} = (I - W)^{-1} X_{new} \beta$. This corresponds to modeling changes in the existing situations. For example, one of the variables in the Boston data set is the average number of rooms per dwelling. Since existing dwellings are often upscaled in size (e.g. the addition of a double storey), the values of the explanatory variables will change. However, in contrast to ordinary regression, in the spatial case there is evidence of real spatial spillovers that need to be estimated: if a few houses in a neighborhood are upscaled, this may result in a price rise for even those nearby dwellings that do not upscale. Thus, it is necessary to predict not just a price of a new $x_{i}$ with given explanatory variables, but also how this perturbation to the existing system could in turn change the existing prices, given an exisiting structure of spatial dependence in $\mathbf{W}$, and the estimated $\beta$. In the case of the Boston data set, we successfully use this approach to predict new prices and spatial spillover effects by making small perturbations to the existing variables and recomputing the new price vector $y_{new}$ using the learnt $W$ and $\beta$. Figure~\ref{bostonFig3} shows, for example, the predicted change in prices ($y_{predicted} - y_{new}$) by raising the average number of rooms variables for 3 locations. The 3 locations themselves go through the largest price change (red points), but also visible are changes to other houses in the area that belong to a similar price segment (shades of blue). Note that the prices for nearby houses that do not belong to the same price cluster do not change as much as those that do belong to the same price cluster. Thus, the prediction of the spatial spillover effect is filtered not just by local spatial distance or connectivity, but by the structure of both short range and long range dependencies learnt in $W$.


\section{Related Work}
\label{relatedwork}
We will review the relevant work in spatial autoregressive models (SAR) and
Alternating Direction Method of Multipliers (ADMM). SAR models were introduced as a generalization of linear regression
to explicitly model spatial autocorrelation in data~\cite{anselin2013spatial,lesage2008introduction}. 
Given a SAR model $ y = \rho\mathbf{W} + \mathbf{X}\beta + \epsilon$,
the parameters $\rho$ and $\beta$ can be estimated using either maximum
likelihood estimation or Bayesian methods. The SAR model is sometimes
referred to as the spatial lag or even the mixed regressive model. 
The computational challenge in the estimation the SAR model arise
due to the need for estimating the spatial dependency parameter $\rho$
as it amounts to optimizing the following function involving a
log determinant term of an $n \times n$ matrix:
\[
\min_{|\rho| < 1} \frac{-2}{n}\ln|\mathbf{I} - \rho\mathbf{W}| + SSE
\]

Since the log determinant term requires the computing the eigenvalues of
$\*W$ several approximation including approximations based on the trace,
chyebyshev's polynomial have been proposed~\cite{kazar2004comparing,pace1997sparse}. In our formulation by making
$\*W$ as the unknown we have avoided the computation of the log det term.
Our ADMM formulation is bottlenecked by the computing the inverse of
$(\*XX^{T} + \*I)$.

%

The ADMM approach is now routinely used to solve problems of the form:~\cite{boyd2011distributed}

\begin{align*}
\mbox{minimize} & f(x) + g(z) \\
\mbox{subject to } & Ax + Bz = c
\end{align*}
ADMM is particularly suitable for solving optimization problems arising in
data mining and machine learning where $f(x)$ represents the data fitting
term and $g(x)$ enforces regularization. While ADMM has not been used
the formulation of the SAR problem, there have been two strands of
work which bear some similarity. In KDD 2015, Hallac et. al.~\cite{hallac2015network}  introduced
the problem of {\em Network Lasso}. Given a graph $G=(V,E)$ where where
each node is associated with a variable $x_{i}$, the network lasso
optimization formulation is
\begin{align*}
\mbox{minimize} & \sum_{i \in V}f_{i}(x_{i}) + \lambda\sum_{(j,k) \in E}w_{jk}\|x_{j} - x_{k}\|_{2}
\end{align*}
Here $w_{jk}$ are elements of the contiguity matrix that is assumed to be given a priori.
The paper tests its approach on California housing
data set and infers clusters based on the price variables by fixing the number of nearest neighbors for each data point. Another strand of research is related to spectral unmixing (SU)~\cite{AmmanouilFR14} where the objective is to learn the so called abudnance matrix $\*W$
given by the equation $\* Y = \*\Phi\*W + \*E$. Here $W$ captures matrix
regression parameters. Similar formulations to
address image denoising problem have been addressed~\cite{yang2013efficient}. 

\section{Conclusion}
\label{conclusions}
We have presented a convex optimization formulation of the spatial
autoregressive regression (SAR) model ($y = \*Wy + \*X\beta + \epsilon$)  
where both the contiguity matrix
$\*W$ and the regression parameters $\beta$ are treated as unknown.
To the best of our knowledge this is the first spatial regression
formulation where the contiguity matrix is inferred from data. We have
developed an ADMM based algorithm to solve the convex optimization problem
which is nearly two orders of magnitude faster than using generic 
convex optimization solvers like CVX. One of the side effects of our
approach is that the inference of $\*W$ gives us a natural way to
cluster the underlying spatial data. Infact our approach can be seen
as a method of simultaneously carrying out regression and clustering on data
while obtaining global regression parameters!
We apply our approach on two housing data sets (Boston and Sydney) where 
the focus is on price prediction. Analysis of the inferred contiguity
matrix reveals the existence of housing submarkets which have their
own internal characteristics and the both short and long range interactions 
between them. We conjecture
that our approach will be of invaluable help to both urban policy makers
and real estate professionals who want to get a deeper insights about
the endogenous aspects of housing markets.



\begin{acknowledgments}
The work in this paper has been supported by a University of Sydney Henry Halloran Trust Grant that funds the Urban Lab @ Sydney and the Housing Lab Incubator. The authors thank Core Logic / R P Data and SIRCA for providing the aggregate level data for Sydney.

\end{acknowledgments}

\end{document}